\documentclass[useAMS,usenatbib]{mn2e}
\usepackage{epsf}
\usepackage{amssymb}
\usepackage{epsfig}
\usepackage[usenames]{color}
\newcommand{\be}{\begin{eqnarray}}

\def\plotone#1{\centering \leavevmode
\epsfxsize=\columnwidth \epsfbox{#1}}

\title[Accretion disks in UCXBs]{X-ray diagnostics of chemical composition of the accretion disk and donor star in ultra-compact X-ray binaries}

\author[F.Koliopanos,  M.Gilfanov and L.Bildsten]{Filippos Koliopanos$^{1}$\thanks{filippos@mpa-garching.mpg.de}, Marat Gilfanov$^{1,2}$ and Lars Bildsten$^3$  \\
$^{1}$MPI f\"ur Astrophysik, Karl-Schwarzschild str. 1, Garching, 85741, Germany\\
$^{2}$Space Research Institute of Russian Academy of Sciences, Profsoyuznaya 84/32, 117997 Moscow, Russia\\
$^3$Kavli Institute for Theoretical Physics, University of California, Santa Barbara, CA 93106-4030, USA}

\begin{document}

\date{Accepted 2013 March 26.  Received 2013 March 26; in original form 2013 February 8}

\pagerange{\pageref{firstpage}--\pageref{lastpage}} \pubyear{2012}

\maketitle

\label{firstpage}

\begin{abstract}
{Non-solar composition of the donor star in ultra-compact X-ray binaries may have a pronounced effect on the fluorescent lines appearing in their spectra due to reprocessing of primary radiation by the accretion disk and the  white dwarf surface.
We show that  the most dramatic and easily observable consequence of the anomalous C/O abundance,  is the significant, by more than an order of magnitude,  attenuation of the K$_\alpha$ line of iron. It is caused by screening of  the presence of iron by oxygen -- in the C/O dominated material the main interaction process for a $E\approx 7$ keV photon is  absorption by oxygen rather than by iron, contrary to the solar composition case. Ionization of oxygen at high mass accretion rates adds a luminosity dependence to this behavior -- the iron line is significantly suppressed only at low luminosity,  $\log(L\rm_X)\la 37-37.5$, and should recover its nominal strength at higher luminosity. The increase of the EW of the $\rm K\rm_\alpha$ lines of carbon and oxygen, on the other hand,  saturates at rather moderate values. Screening by He is less important, due to its low ionization threshold and because in the accretion disk it is mostly ionized. Consequently, in the case of the He-rich donor,  the iron line strength remains close to its nominal value, determined by the iron abundance in the accretion disk.  This opens the possibility of constraining the nature of donor stars in UCXBs by means of X-ray spectroscopy with moderate energy resolution.
}
\end{abstract}

\begin{keywords}
accretion, accretion discs -- line: formation -- line: profiles -- X-rays: binaries 
\end{keywords}

\section{Introduction}

Ultra compact X-ray binaries (UCXBs) are a sub-group of X-ray binaries with orbital periods of less than 1 hour. Their small orbital periods do not allow for a hydrogen rich, main sequence donor (e.g. \citealt{1984ApJ...283..232R}; \citealt*{1986ApJ...311..226N}). The most likely scenarios for their formation \citep*[for details see e.g.][]{1993ARep...37..411T, 1995ApJS..100..233I, 2005ASPC..330...27N} predict that the donor star in such systems is a white dwarf (WD)  or a helium star.  Driven by the loss of the orbital angular momentum due to gravitational wave radiation, UCXBs are often observed as persistent and relatively luminous X-ray sources with luminosities in the  $10^{36} - 10^{38}{\rm erg\,s^{-1}}$ range \citep{1986ApJ...311..226N,2004ApJ...607L.119B}.

Given the nature of the donor star, the accreting material in UCXBs should have a chemical composition consistent with the ashes of H burning (mostly He and $\rm ^{14}N$), He burning (mostly C/O) or carbon burning (mostly O/Ne). Depending on the binary's formation channel, it can vary from C/O-rich to He-rich. Indeed, optical observations of several  UCXBs, for example, 4U 0614+091, 4U 1543-624 and 2S 0918-549 suggest accretion of C/O-rich material (\citealt{2004MNRAS.348L...7N}; \citealt*{2006MNRAS.370..255N}; \citealt{2006A&A...450..725W}). On the other hand,  in the case of 4U 1916-05  they reveal evidence pointing to a He-rich donor \citep{2006MNRAS.370..255N}.  Modeling of type I X-ray bursts from 4U 1820-30 suggest that the accreting material in this system is also helium dominated \citep{1995ApJ...438..852B,2003ApJ...595.1077C}.

X-ray spectra of X-ray binaries usually contain the so called reflected component \citep[e.g.][and references therein]{2010LNP...794...17G}. This component is produced due to reprocessing of primary emission  by  the optically thick Shakura-Sunyaev  accretion disk and by the surface of the donor star facing the compact object. The primary emission may originate in a hot optically thin corona, in the accretion disk itself or, in the case of a NS accretor, in the boundary layer on the surface of the star, and carries  most of the energy. Depending on its origin, the spectrum of the primary emission may vary from soft thermal  to  hard power law-like spectrum. Although the reflected component is energetically insignificant, it carries information about the geometry of the accretion flow \citep*[e.g.][]{1999A&A...352..182G} and, via fluorescent lines and absorption edges of metals, about the chemical composition, ionization state and kinematics of the accretion disk material.

Reprocessing of X-ray radiation by the accretion disk and by the surface of the donor star has been studied extensively by many authors, starting from the seminal paper by \cite*{1974A&A....31..249B}. The shape and strength of the iron $\rm K\rm_{\alpha}$ fluorescent line has been  investigated by \cite{1978ApJ...223..268B} and \cite{1979SoPh...62..113B}. Semi-analytical expressions for reflection spectra including the effects of both photoionization and Compton scattering, have been derived by \cite{1988ApJ...335...57L} and \cite*{1988ApJ...331..939W}, and formulated  in terms of K-shell fluorescence and the characteristic Compton hump between ~10\,keV and ~300\,keV. Later on, the effect of ionization of the accreting material on reflected spectra has been included (e.g. \citealt*{1993MNRAS.261...74R}; \citealt{1994ApJ...437..597Z}; \citealt*{2000ApJ...537..833N}). The need for better accuracy and more realistic and complex geometries has led to application of Monte-Carlo (MC) methods that complimented and enhanced analytical calculations. Many authors have computed detailed models based on Monte-Carlo techniques \citep*[e.g.][]{1991MNRAS.249..352G,1993MNRAS.262..179M,2001MNRAS.327...10B}.

Despite of the amount of effort invested in studying reflection of X-ray emission from optically thick media, all prior work concentrated on the  $\sim$ solar abundance case, with only moderate variations of the element abundances considered in some of the papers.
On the other hand, in the case of UCXBs, we expect that accreting material may have significantly non-solar abundances, for example with all hydrogen and helium being converted to carbon and oxygen. Such drastic abundance modifications should  lead to strong changes in the properties of the reflected spectrum, especially in its fluorescent line content.
This problem is investigated in the present paper.  The composition of the accreting material is discussed in Section 2. In Sections 3--4, we consider an idealized case of an optically thick slab of neutral material in order to identify main trends and then (Section 5) discuss modifications to this picture which may be introduced by gravitational settling of heavy elements in the white dwarf envelope and ionization of the accretion disk material by viscous heating and irradiation. We use simple analytical calculations (Section 3 and Appendix) to illustrate the physical origin of the  main dependencies  and then utilize  the Monte-Carlo technique to compute reflected spectra for strongly non-solar abundances of the type expected in the donor stars in UCXBs (Section 4). We mainly focus on the strengths of fluorescent lines of the  elements expected to be abundant in different formation scenarios of UCXBs.

\begin{table}
 \caption{Abundances of elements for different types of white dwarfs used throughout the paper.}
 \label{tab:abund}
 \begin{tabular}{@{}lccc}
  \hline
   Element &He  & C/O &O/Ne       \\
   &{Pandey} & Garc\'{\i}a Berro & Gil-Pons \&,  \\
        & et.al.  &  et.al. &Garc\'{\i}a Berro \\

 \hline
H&  {\it1.99}${\it\cdot {10^{-6}}}$& -&        -\\
He& {\it0.997}&    -&        -\\
C&  {\it1.58}${\it\cdot {10^{-3}}}$& {\it0.563}&    -\\
O&  {\it3.97}${\it\cdot {10^{-4}}}$& {\it0.422}&    {\it0.649}\\
Ne& 3.66$\cdot {10^{-4}}$& {\it1.37}$ {\it\cdot {\it{10^{-2}}}}$&   {\it0.262}\\
Na& 6.06$\cdot {10^{-6}}$& 2.08$\cdot {10^{-5}}$& {\it4.93}${\it\cdot {10^{-2}}}$\\
Mg& 1.08$\cdot {10^{-4}}$& 3.69$\cdot {10^{-4}}$& {\it3.88}${\it\cdot {10^{-2}}}$\\
Si& 1.01$\cdot {10^{-4}}$& 3.44$\cdot {10^{-4}}$& 4.43$\cdot {10^{-4}}$\\
S&  4.59$\cdot {10^{-5}}$& 1.57$\cdot {10^{-4}}$& 2.02$\cdot {10^{-4}}$\\
Ar& 1.27$\cdot {10^{-5}}$& 4.34$\cdot {10^{-5}}$& 5.58$\cdot {10^{-5}}$\\
Cr& 1.33$\cdot {10^{-6}}$& 4.54$\cdot {10^{-6}}$& 5.84$\cdot {10^{-6}}$\\
Mn& 6.94$\cdot {10^{-7}}$& 2.38$\cdot {10^{-6}}$& 3.06$\cdot {10^{-6}}$\\
Fe& 9.18$\cdot {10^{-5}}$& 3.14$\cdot {10^{-4}}$& 4.04$\cdot {10^{-4}}$\\
Co& 2.36$\cdot {10^{-7}}$& 8.07$\cdot {10^{-7}}$& 1.04$\cdot {10^{-6}}$\\
Ni& 5.04$\cdot {10^{-6}}$& 1.73$\cdot {10^{-5}}$& 2.22$\cdot {10^{-5}}$\\
Cu& 4.59$\cdot {10^{-8}}$& 1.57$\cdot {10^{-7}}$& 2.02$\cdot {10^{-7}}$\\
Zn& 1.13$\cdot {10^{-7}}$& 3.86$\cdot {10^{-7}}$& 4.96$\cdot {10^{-7}}$\\

  \hline
 \end{tabular}

 \medskip
{Abundances are by number of particles. The references to the original abundance calculations are given in the column titles. The numbers written in italics are from these calculations. Abundances of other elements were fixed at the solar values in mass units and then converted to concentration abundances.}
\end{table}

\begin{figure*}
\includegraphics[width=0.48\textwidth]{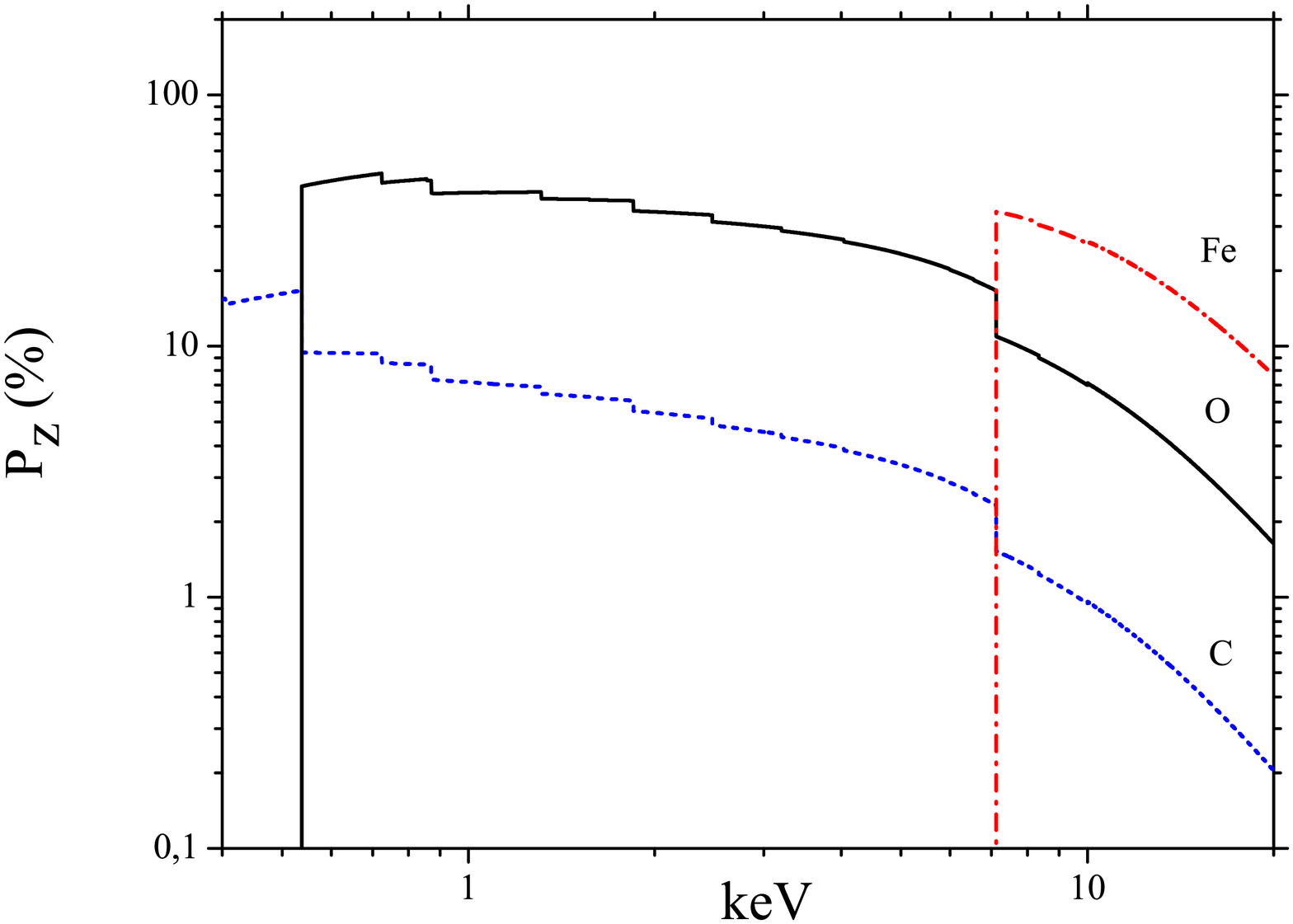}
\includegraphics[width=0.48\textwidth]{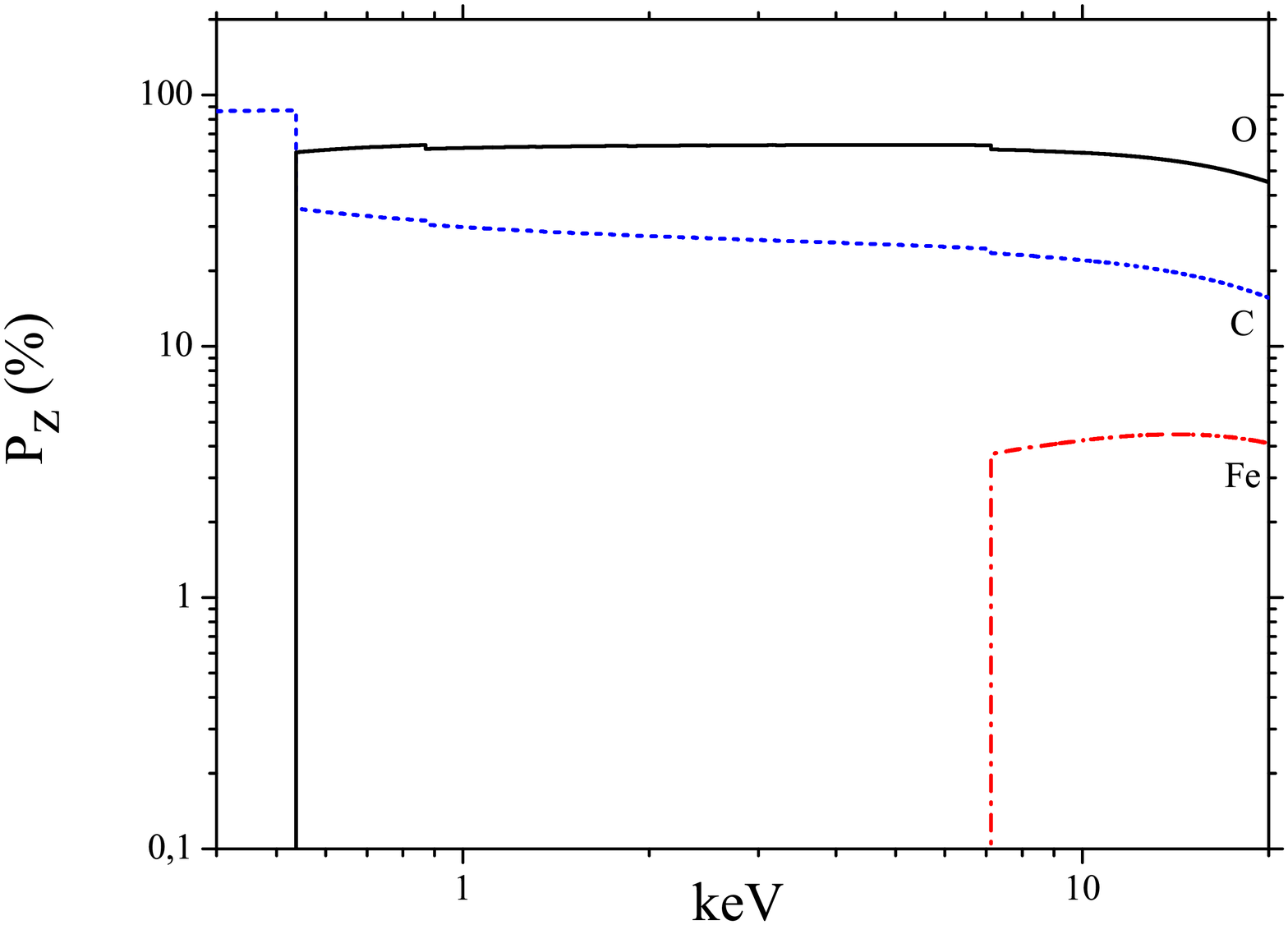}
\caption{The probability $P\rm_Z(E)$ (eq.(\ref{eq:pz})) for an incident photon to be absorbed due to K-shell ionization of  carbon (blue dashed line), oxygen (black solid line) and iron (red dash-dotted line), in the solar abundance case (left panel) and for the composition  of a C/O white dwarf as tabulated in Table \ref{tab:abund} (right panel).}
\label{fig:pz_vs_e}
\end{figure*}

\section{Composition of the accreting material}

Different initial parameters and the environment of  UCXB progenitors may lead to a variety of donors \citep*[e.g.][]{1986A&A...155...51S,2002ApJ...565.1107P,2002A&A...388..546Y} -- non-degenerate He star, He white dwarf, C-O or O-Ne-Mg white dwarf. As it will become clear later, from the point of view of classification of the reflected spectra,  the variety of abundance patterns can be broadly divided into two types -- (i) He-rich and (ii) C/O/Ne/Mg-rich.

A typical isolated C/O white dwarf is expected to consist of a core mostly  made of a mixture of  carbon and oxygen,  surrounded by a He-rich layer of up to $ \rm {10^{ - 2}}~{M_ \odot }$. On top of the helium layer there may be  a thin layer of hydrogen of up to $ \rm {10^{ - 4}}~{M_ \odot }$  \cite[e.g.][]{1995stre.conf....1K,2010A&ARv..18..471A}. A late shell flash could produce a C/O-rich envelope comprised of $\sim 30\%$ of  He from convective shell burning \citep{1983ApJ...275L..65I}. The mass of the He-C-O mantle surrounding the C/O core can grow up to $ \rm {10^{ - 1}}{M_ \odot }$, itself enveloped by a thin layer of H in the case of a hybrid white dwarf \citep{1985ApJS...58..661I,1987ApJ...313..727I}. On the other hand, if the  initial mass of the companion star was in the  $\rm \sim 8-11{M_ \odot }$  range,  a UCXB with an O-Ne white dwarf donor may be formed \citep{2001A&A...375...87G}.

In the case of a white dwarf in a binary system, this basic structure will be modified by a co-evolution with the companion star. H and He layers can either be stripped away during the initial stages of binary interaction \citep*{2012ApJ...758...64K} or be gradually depleted due to accretion. Indeed, typical luminosities of UCXBs are in the $\sim 10^{36} - 10^{38}$ erg/s range, implying the mass accretion rate in the $\sim 10^{-10}-10^{-8}$ M$_\odot$/yr range. At this rate, a surface layer of $\la10^{-2}$ M$_{\odot}$ will be depleted within $\la 1-100$ Myrs, which is (much) shorter than the expected life times of such systems.

If the white dwarf donor has been completely stripped of its H and He layers, the chemical composition of the accreting material will be determined by its  core. For the purpose of this calculation we will ignore the complexity of the possible abundance patterns and assume the following mass fractions:  $\rm C = O \simeq 0.49$ and $\rm Ne \simeq 0.02$ \citep[e.g.][]{2008ApJ...677..473G}. More massive white dwarfs are expected to have cores composed mainly of oxygen and neon \citep*[e.g.][]{1996ApJ...460..489R}, in which case we assume the following composition: $\rm Ne \simeq 0.28$, $\rm O \simeq 0.55$, $\rm Mg \simeq 0.05$, $\rm Na \simeq 0.06$ \citep{2001A&A...375...87G}.

The He-star and evolved secondary donor scenarios yield an accreting material that is He-dominated, perhaps with some He burning products and traces of H depending on the phase at which the evolved secondary star started its Roche lobe overflow. \cite{2001MNRAS.324..937P} have performed detailed calculations for the abundances of cool He-stars and obtained the following values: He=0.99, C=0.0052, N=0.0016, O=0.0018. For an evolved main sequence star companion \citet{2010MNRAS.401.1347N} used the Eggleton stellar evolution code {\small{TWIN}} to model the abundance pattern at the donor's surface. They predict material that is He-rich with less than 0.01 O and N and traces of C.

For the purpose of this paper we will assume that the mass fractions of all other elements, not included in the above calculations, are equal to their solar values. This is equivalent to the assumption that the total masses of these elements in the star  did not change in the course of its evolution towards a white dwarf. The corresponding mass fractions were then converted to the concentration abundances taking into account changed particle concentrations of H, He, C, O etc.  Solar abundances for elements with Z=1-30 were adopted from \citet{1992PhyS...46..202F}, elements not listed in this tabulation were taken from \citet{1998SSRv...85..161G}.
The abundance patterns for different types of white dwarfs used throughout the paper are summarized in Table~\ref{tab:abund}. In the Table, the values obtained in the white dwarf abundance calculations described above are given in italic,  abundances of elements which mass fractions were fixed at their solar value are shown in roman font.

\section{Qualitative picture }

In this section and in section 4, we consider an idealized case of reflection from cold and neutral material. Although, at a sufficiently low mass accretion rate, this is a reasonable approximation for carbon and oxygen, it breaks down for helium in the accretion disk   at any luminosity relevant to UCXBs. In the case of a cold white dwarf, it may hold  on its surface.  The effect of ionization is discussed in detail in Section 5.

Although the strengths of fluorescent lines depend on the number of parameters, such as spectral index of the primary radiation, incidence angle, ionization state etc., \citep*[e.g.][]{1991MNRAS.249..352G,2002MNRAS.329L..67B} variation of these parameters within their plausible ranges leads to  rather moderate changes in the equivalent widths.
Much more dramatically  the EWs of fluorescent lines  are affected by changes in the relative abundances of different elements. As it turns out, an increase in the abundances of carbon and oxygen, not only changes the strengths of fluorescent  lines of these elements, but has a much easier detectable  effect on the iron line. Indeed, as it will be shown below, for the typical X-ray spectra of UCXBs,  the equivalent widths of carbon and oxygen lines increase by $\sim 2-10$ times with respect to their solar abundance values, but still remain in the hardly detectable  sub-eV range. On the other hand, the iron line drops  more than 10-fold, from easily observable $\sim 100$ eV to the $\sim$ few eV range.

The effect of elemental abundances on the strength of emission lines, can be illustrated by the following simple calculation.  Let's consider an optically thin layer of material illuminated by a photon beam traveling along the axis normal to its surface.  The probability that an incident photon with energy $E$ is absorbed due to K-shell ionization of a  given element Z (rather than being absorbed by other elements or scattered by electrons)  is given by the following expression:
\begin{equation}
{P\rm_Z}(E) = \frac{{{A\rm_Z}\,{\sigma \rm_{K,Z}}(E)}}{{{A\rm_Z}\left [{\sigma \rm_{\rm{Z}}}(E) + Z\,{\sigma \rm_{KN}}(E)\right ] + \sum\limits_{Z' \ne Z} {\sigma '(E)} }}
\label{eq:pz}
\end{equation}
where ${\sigma \rm_Z}$ -- the total absorption cross section of element Z, ${ \rm \sigma \rm_{K,Z}}$ is its K-shell absorption cross section, $ \rm {A\rm_Z}$ -- the  abundance of element Z by number, $\rm \sigma \rm_{KN}$ -- the Klein-Nishina cross section.  For compactness of the formula we denote $\rm \sigma'(E) = {A\rm_{Z'}}[{\sigma \rm_{Z'}}(E) + Z'{\sigma \rm_{KN}}(E)]$ -- the total cross section due to element $\rm Z'$, including Thomson scattering on its electrons (see the Appendix for further details). Obviously, the quantity $P\rm_Z(E)$ determines what fraction of incident photons with energy E will contribute to the production of the fluorescent line of the element $Z$ and offers a simple way to qualitatively investigate the dependence of the EW of the line on element abundances. Moreover, as it will be shown later, a modification of eq.(\ref{eq:pz}) gives a reasonably accurate method to analytically compute EWs of fluorescent lines.

\begin{figure}
\plotone{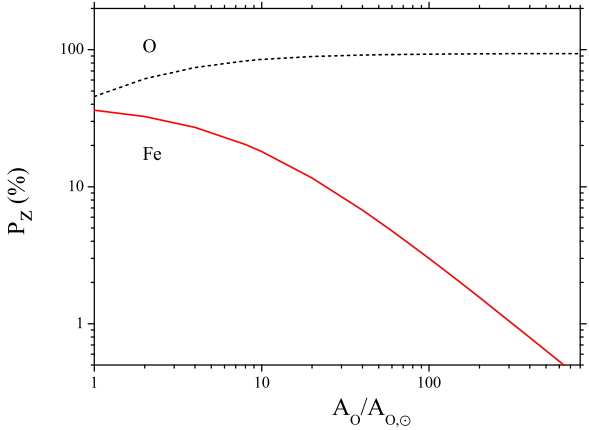}
\caption{ The probability of absorption by oxygen and iron calculated at the respective K-edges, versus oxygen abundance $A\rm_O$. The latter is expressed in solar units.
\label{fig:pz_vs_ab}}
\end{figure}

In Fig. \ref{fig:pz_vs_e}  we plot the probability $P\rm_Z(E)$ for carbon, oxygen, and iron versus energy. The left panel  was computed for an atmosphere with solar abundances, the right panel --  for the abundances appropriate for a C/O white dwarf (Table \ref{tab:abund}).  As one can see from the plot, in a solar abundance  case, oxygen is the dominant absorbing element for photon energies of up to $\approx 7$ keV. However, above the Fe K-shell ionization threshold of 7.11 keV, the value of $ \rm P\rm_{Fe}$ jumps and exceeds that of $ \rm P\rm_{O}$, so that at higher energies the majority of photons are absorbed by iron and contribute to its fluorescent line. For the chemical composition of a C/O WD, however, the picture changes dramatically. Even though the K-shell photo-absorption cross-section for iron is larger than the oxygen cross-section at these energies, the increased abundance of oxygen makes it the main absorbing agent in the entire energy range. As a result,  the $ \rm K \alpha$ line of iron  will be significantly  suppressed.

As oxygen abundance increases, $ P\rm_O$ will increase linearly with $A\rm_O$, insofar its contribution to the denominator in eq.(\ref{eq:pz})  remains relatively small. However, at sufficiently large abundances, the oxygen term prevails and $ P\rm_O$ saturates at a value determined by the ratios of cross-sections, $P\rm_O\sim \sigma\rm_{K,Z}/(\sigma\rm_{Z}+\sigma\rm_{KN})$. On the contrary,  $ P\rm_{Fe}$ will continue to decrease due to unlimited increase of $\rm \sigma'$ in the denominator in eq.(\ref{eq:pz}). This behavior is illustrated in Fig. \ref{fig:pz_vs_ab}\footnote{ Note that for the purpose of this plot we fixed abundances of all elements except oxygen at their solar value, i.e. no condition of the  nucleon number conservation was imposed.
Such an abundance sequence does not represent any of the WD compositions and is employed here to investigate the behavior of equivalent widths in the limit of high A$_{\rm O}$ (or, equivalently, high O/Fe). If we use the O/Fe ratios ($\rm (O/Fe)_\odot\approx 26$, $\rm (O/Fe)_{CO}\approx 1.3\cdot 10^3$) to characterize the oxygen overabundance, a C/O white dwarf would approximately correspond to the $ \rm A_O/A_{O,\odot}\sim 50$. }
 where we plot the values of $ P\rm_O$ and $P\rm_{Fe}$ estimated at their respective K-edges, versus $ A\rm_O$.  As is evident from the plot, $ P\rm_O$ increases with $ A\rm_O$ until the latter reaches a value of $\sim 20-25$ times  solar value. At this abundance, oxygen completely dominates the opacity and nearly all incident photons that are not scattered on electrons, will be absorbed by the oxygen, regardless of their energy. Further increase of oxygen abundance does not lead to an increase of it fluorescent line strength.  On the other hand, the ${P}\rm_{Fe}$ curve shows unlimited  decreases as $A\rm_O$ increases, asymptotically $P\rm_{Fe}\propto A\rm_O^{-1}$.

\begin{figure*}
\includegraphics[width=0.48\textwidth]{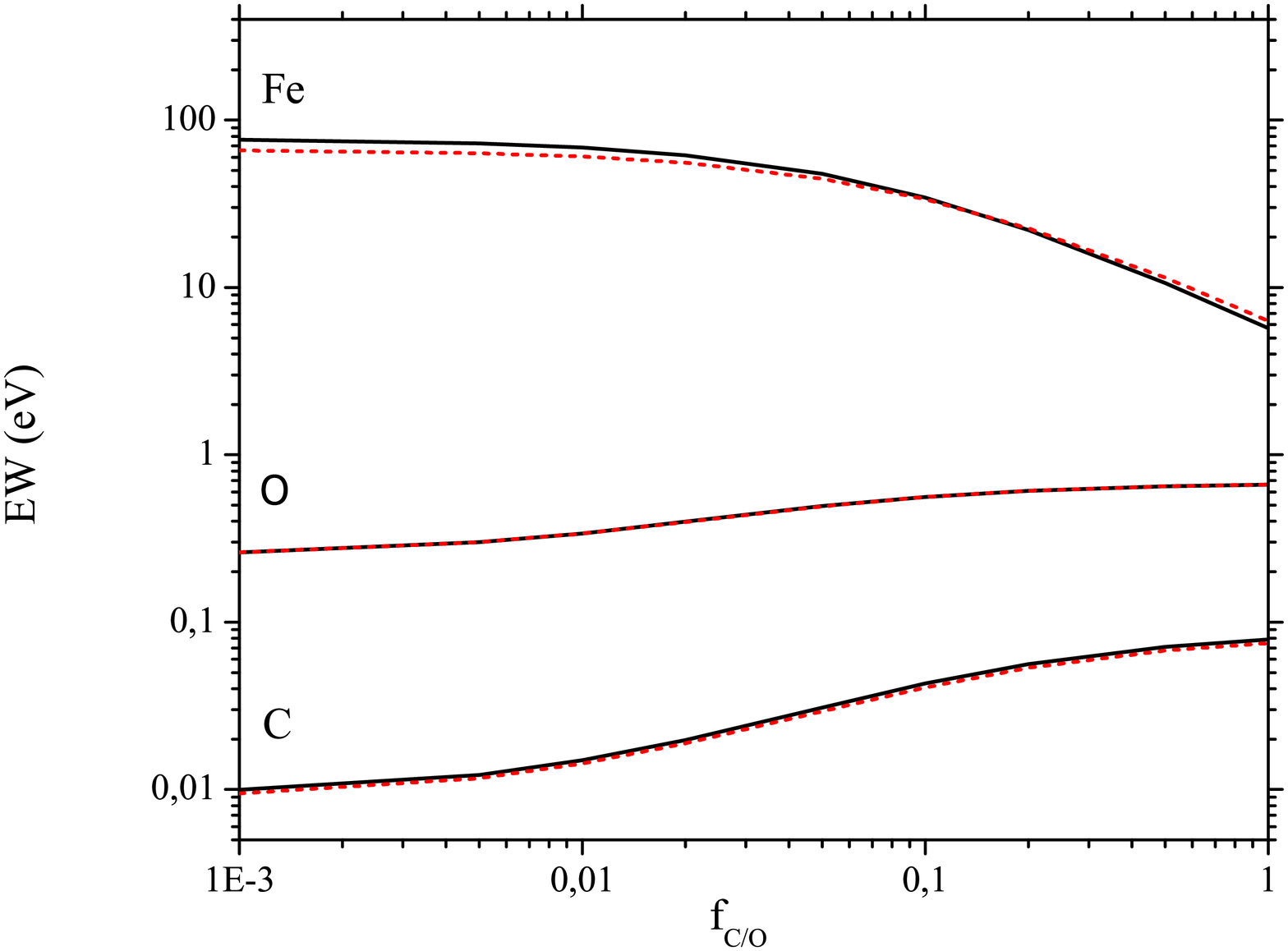}
\includegraphics[width=0.48\textwidth]{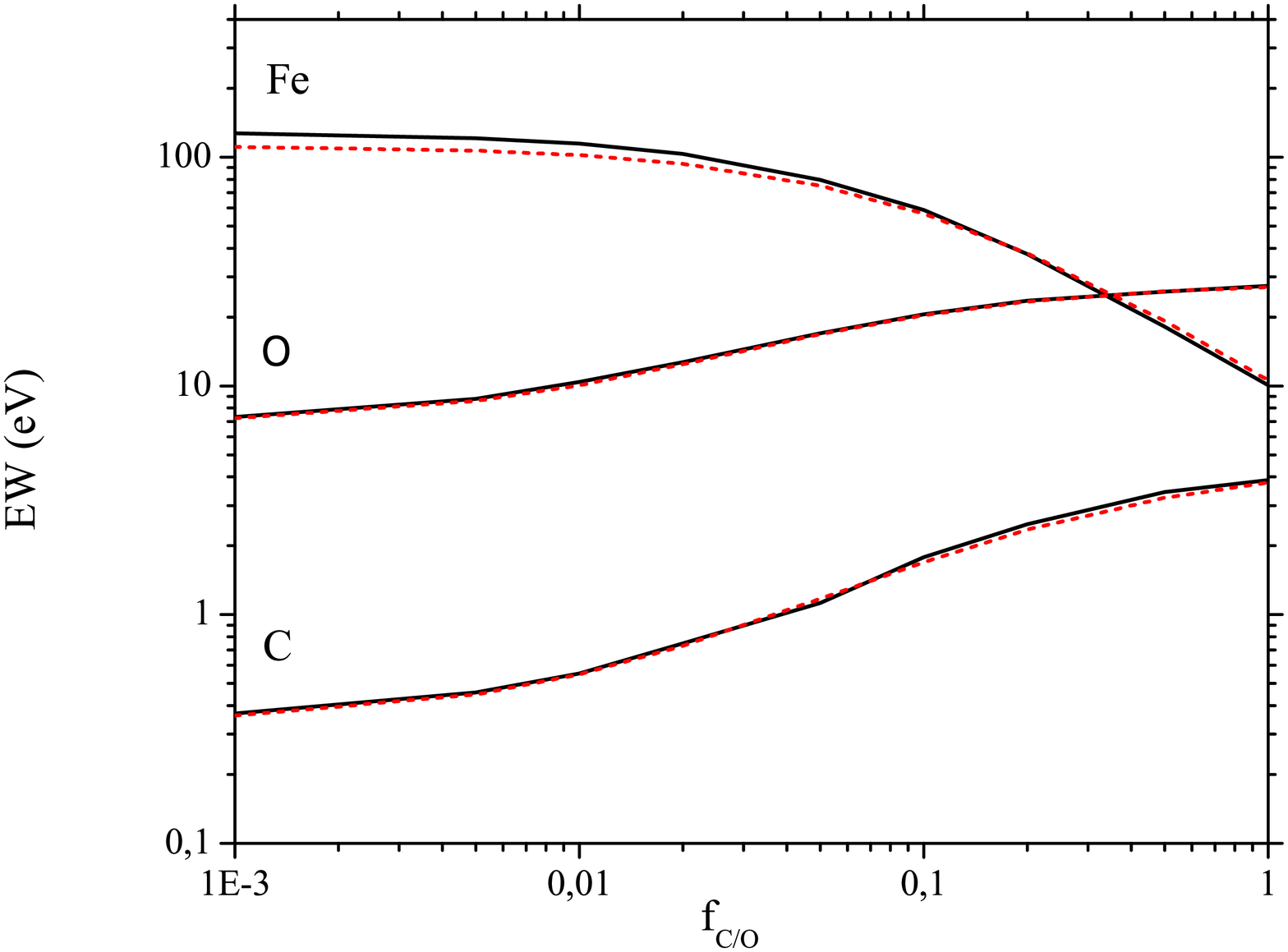}
\caption{ The equivalent widths of $ \rm K\alpha $ lines of C, O and Fe  plotted against ${f\rm_{C/O}}$, the fraction of H and He "converted" to C and O (section \ref{sec:co}, eq.\ref{eq:fco}). The left panel shows results for a power law  incident spectrum   with photon index of $\rm \Gamma=1.9$  and the right panel -- for black body radiation with kT=2.5 KeV. Black solid lines show results of Monte-Carlo calculations, the dashed lines (red in the color version) were computed in the single scattering approximation as described in the Appendix.
}
\label{fig:co}
\end{figure*}

\section{Results}

\subsection{Method of calculation}

We consider plane-parallel, semi-infinite  geometry.  The disk material is assumed to be cold and neutral.
We use the Monte Carlo (MC) technique following the prescription by \citet*{1983ASPRv...2..189P}.  We account for $\rm K_\alpha $ and $\rm K_\beta $ lines for all elements from Z=3 to 30 and do not distinguish between the $\rm K_{\alpha_1} $ and $\rm K_{\alpha_2} $ lines. Photoionization cross sections for given abundances are calculated using the analytical fits for partial photoionization cross-sections from \citet{1996ApJ...465..487V}, the scattering cross-section was described by the Klein--Nishina formula and fluorescence yields for neutral elements from Li to Zn are from \cite{1972RvMP...44..716B}. The incident  spectrum can be modeled either as a beam at a specific angle with respect to the normal to the surface or an isotropic point source above the disk surface. The output reflected emission can be registered at a specific angle or integrated over all viewing angles. In computing the equivalent widths of lines with respect to the total emission the reflected emission was mixed with the primary radiation as described in the Appendix.

X-ray binaries are known to have two distinct spectral states: a bright high/soft state and a less luminous low/hard state \citep[e.g.][]{2010LNP...794...17G}. In the hard spectral state, Comptonization of soft photons on hot electrons is the most likely mechanism for the creation of the hard spectral component, which, in the energy range of interest, usually has a power law shape. In the soft state, the primary emission may originate in the accretion disk itself or in the boundary layer on the surface of the neutron star \citep*[e.g.][]{1986SvAL...12..117S,1999AstL...25..269I,2001ApJ...547..355P}. Correspondingly, the incident spectrum is represented by either a power law in which case it  is characterized  by a photon index $\Gamma$, or  by a thermal component originating in the NS boundary layer. In the latter case, we approximated the boundary layer radiation with a black body spectrum with kT = 2.5 keV \citep*{2003A&A...410..217G}.

Along with Monte Carlo simulations, we also performed simple analytical calculations in the single scattering approximation. The formulae are derived in the Appendix and can be used for quick analytical estimates of the strengths of the fluorescent lines for different abundance patterns.

\subsection{Reflection from C/O-rich material}
\label{sec:co}

In order to investigate the dependence of the line strengths on the C/O abundance we consider a sequence of chemical compositions with increasing C/O abundance.  Abundances of H and He were reduced along the sequence so that the total number of nucleons was conserved. Abundances of other elements (by mass) remained fixed at their solar values, as well as the abundance ratio of carbon and oxygen. The position along the sequence is defined by the parameter
\begin{equation}
f_{\rm{C/O}}=\frac{n_{\rm{C}}^{ } m_{\rm{C}}^{ } +n_{\rm{O}}^{ }  m_{\rm{O}}^{ } }{n_{\rm{H}}^{ } m_{\rm{H}}^{ }+n_{\rm{He}}^{ } m_{\rm{He}}^{ }+n_{\rm{C}}^{ } m_{\rm{C}}^{ } +n_{\rm{O}}^{ }  m_{\rm{O}}^{ } }
\label{eq:fco}
\end{equation}
where $n_i$ is  concentration of the element $i$ and $m_i$ is its atomic weight. Obviously, $f\rm_{C/O}$ has the meaning of the fraction of hydrogen and helium "converted" to carbon and oxygen and changes from 0 to 1.
We performed calculations of the reflected spectrum and computed the equivalent widths of fluorescent lines on a grid of values of ${f\rm_{C/O}}$ in this range.  The calculations were performed with Monte-Carlo simulations and  using analytical expressions from the Appendix. The incident photons were assumed to illuminate the disk surface isotropically and the reflected emission was integrated over all outgoing directions. The results are shown in  Fig. \ref{fig:co} for carbon, oxygen and iron $\rm K_\alpha$ lines, for both types of the primary emission spectrum.  The plots show the equivalent widths of lines versus  ${f\rm_{C/O}}$.

The overall behavior of equivalent widths  is similar for both types of incident spectrum. It is mostly determined by $P\rm_Z$ (Fig. \ref{fig:pz_vs_e}) and, as expected, the curves have shapes similar to those in Fig. \ref{fig:pz_vs_ab}. Initially, the EWs of carbon and oxygen lines increase nearly in direct proportion to their abundance, however they quickly  saturate at ${f\rm_{C/O}}\sim20\%$. This corresponds to the oxygen overabundance of $ \rm A\rm_O/A_{O,\odot}\sim 10$ in Fig. \ref{fig:pz_vs_ab}. The initial increase of the oxygen line EW is somewhat milder than that of carbon due to screening of oxygen by carbon itself. In the case of the power law incident spectrum, the EWs of carbon and oxygen lines remain at sub-eV level, even for a totally C/O-dominated composition.  For a black body incident spectrum, however,  these lines can become rather prominent at large  C/O-abundances, reaching a value of $\rm\sim 20\,eV$ for oxygen and $\rm\sim 2\, eV$  for carbon. The iron line is the most prominent fluorescent line for solar abundances. Nevertheless, for both types of the incident spectrum,  its EW drops below $\sim 10$ eV for the C/O dominated composition.

\begin{figure*}
\includegraphics[width=0.48\textwidth]{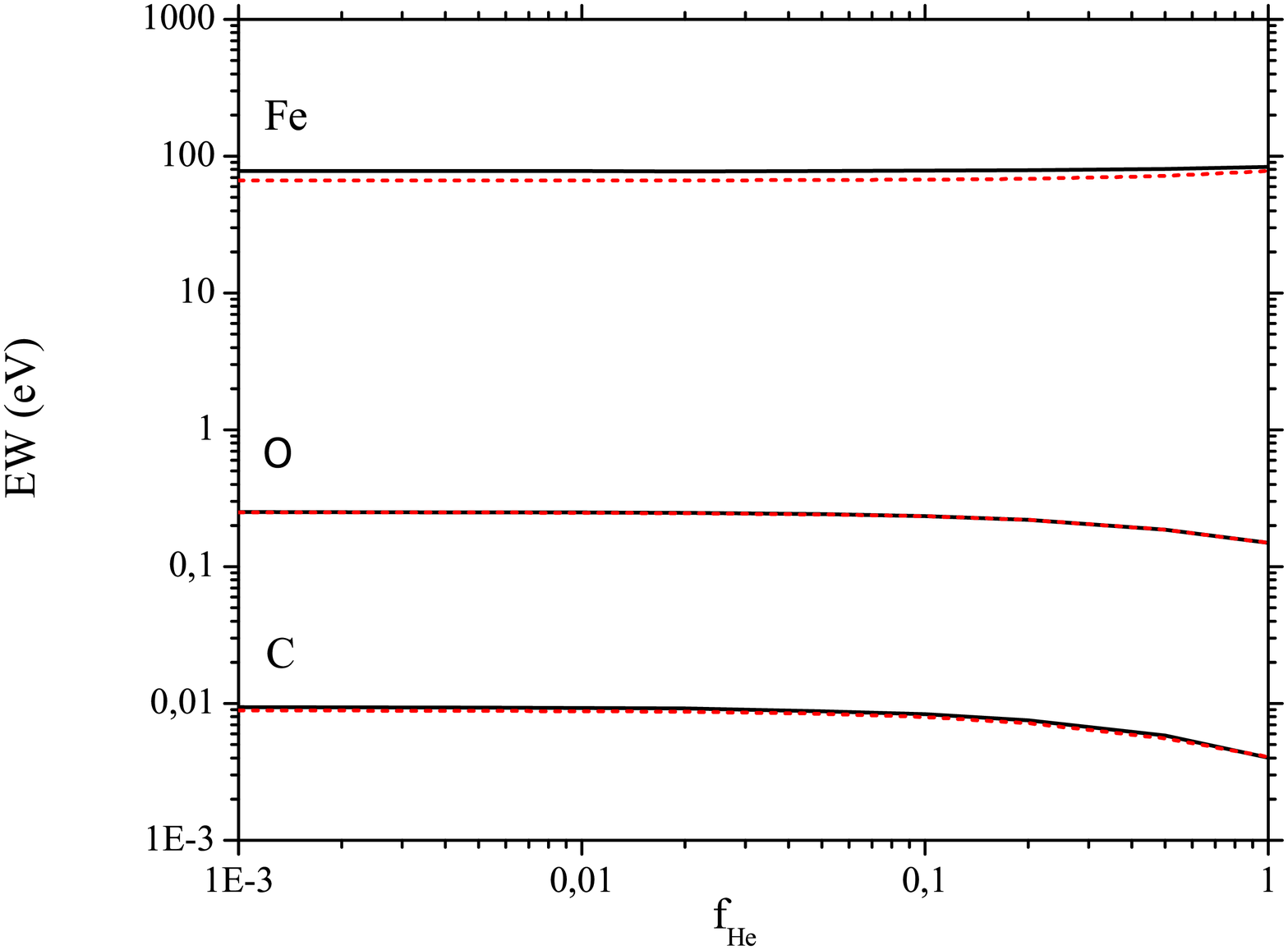}
\includegraphics[width=0.48\textwidth]{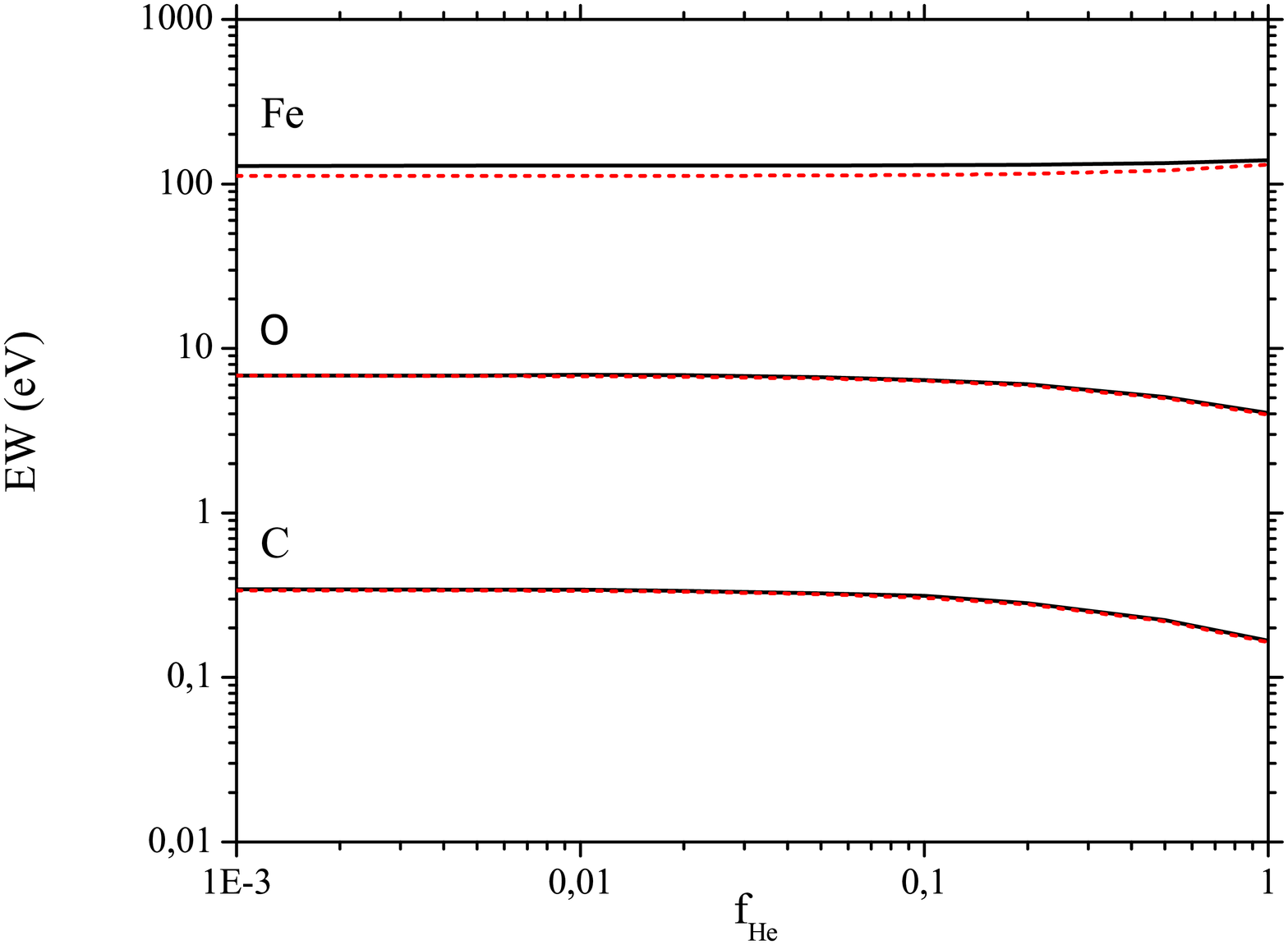}
\caption{The equivalent widths of $ \rm K\alpha $ lines of C, O and Fe  are plotted against ${f\rm_{He}}$, the fraction of H  "converted" to He (cf. eq.\ref{eq:fco}). The left panel shows results for a power law incident spectrum with photon index of $\rm \Gamma=1.9$ and the right panel -- for black body radiation with kT=2.5 keV. Black solid lines show results of Monte-Carlo calculations, the dashed lines (red in the color version) were computed in the single scattering approximation as described in the Appendix.}
\label{fig:he}
\end{figure*}

\subsection{Reflection from He-rich material }

We follow the same approach to investigate the dependence of the fluorescent lines strengths on the helium abundance as we used for C/O. The  results are presented in Fig. \ref{fig:he}, where we plot equivalent widths of carbon, oxygen and iron lines against ${f\rm_{He}}$, defined similarly to $f\rm_{C/O}$ in eq.(\ref{eq:fco}).

Similar to the C/O case, the increased abundance of He would have a screening effect on the elements with the charge $Z>2$, however, the effect is much milder than in the case of C/O dominated material. The reason is the much larger solar abundance of helium and the significant drop of its photoabsorption cross-section  at the energies corresponding to K-edges of  elements which $\rm K_\alpha$ lines are in the X-ray domain. The latter factor is especially important  for iron. Correspondingly, the overall effect is stronger for lines of low-Z elements, such as  carbon and oxygen and is negligible for iron. For example, the equivalent widths of carbon and oxygen lines decrease by $\sim 2$ times in the He-dominated case. In Fig. \ref{fig:he}, one can also notice a slight, $\sim 10\%$, increase of the  EW of the iron relative to its solar value. This is a consequence of the definition of the abundance sequence, and does not have any physical meaning.  It is caused by  the decrease of the number of electrons, as H is being replaced by He along the abundance sequence.

\subsection{Realistic WD compositions}

We now compute equivalent widths of fluorescent lines for realistic abundance patterns expected for different types of white dwarfs.  For  these we used the results of \citet{2008ApJ...677..473G}, \citet{2001A&A...375...87G} and \citet{2001MNRAS.324..937P} as detailed in the Section 2 and summarised in Table \ref{tab:abund}.  The equivalent width of lines were computed with our Monte Carlo code assuming an isotropic point source above the disk surface (lamppost configuration). The reflected spectrum was integrated over all viewing angles. The results of these calculations are presented in the Table \ref{tab:ews}.

The results are qualitatively similar to those obtained in the previous subsections.  Perhaps a new feature is the appearance  of the neon line in the case of the O/Ne white dwarf, which becomes especially prominent in the case of a black body incident spectrum, where its equivalent width can reach   $\sim 15$ eV.

\begin{table*}
 \centering
 \begin{minipage}{140mm}
  \caption{Equivalent width of prominent fluorescent lines for realistic white dwarf compositions. Results of Monte-Carlo calculations. }
\label{tab:ews}
  \begin{tabular}{lcrrrrrrrr}
  \hline
    &  &      &        & Power law & & & & Black Body & \\
  \hline
   Element &Energy& ~~~~~~~Solar & He & C/O & O/Ne & ~~~~~Solar &He & C/O  & O/Ne   \\
       &keV & &   &   & &&  &  &  \\
 \hline
C&	    0.277&	0.01&	$<0.01$&	0.11&	-&	0.35&	0.24&	8.21&	-\\
O&		0.525&	0.25&	0.03&	0.40&	0.72&	7.10&	0.70&	13.8&	31.0\\
Ne&		0.849&	0.19&	0.20&	0.06&	0.76&	3.14&	3.19&	1.00&	14.8\\
Na&		1.041&	0.01&	0.01&	$<0.01$&	0.23&	0.09&	0.10&	$<0.01$&	3.51\\
Mg&		1.254&	0.25&	0.27&	$<0.01$&	0.37&	2.90&	3.17&	0.10&	4.36\\
Si&		1.740&	0.84&	0.94&	0.03&	0.02&	7.23&	8.24&	0.26&	0.14\\
S&		2.307&	1.13&	1.30&	0.04&	0.02&	7.19&	8.36&	0.30&	0.14\\
Ar&		2.957&	0.80&	0.94&	0.04&	0.02&	3.84&	4.50&	0.17&	0.08\\
Ca&		3.690&	1.03&	1.22&	0.05&	0.03&	3.60&	4.30&	0.18&	0.09\\
Cr&		5.415&	0.80&	1.00&	0.05&	0.03&	1.65&	2.14&	0.11&	0.05\\
Mn&		5.899&	0.52&	0.65&	0.04&	0.02&	0.98&	1.24&	0.07&	0.04\\
Fe&		6.397&	76.0&	92.8&	6.58&	3.11&	127&	157&	10.9&	5.17\\
Fe $(\rm K_{\beta})$&	7.058&	12.6&	15.7&	1.17&	0.54&	17.3&	21.6&	1.60&	0.80\\
Ni&		7.470&	5.86&	7.00&	0.68&	0.33&	7.75&	9.37&	0.92&	0.44\\
Ni $(\rm K_{\beta})$&	8.265&	0.93&	1.11&	0.12&	0.06&	1.04&	1.28&	0.13&	0.06\\

\hline
\end{tabular}
\medskip

{Equivalent widths are in eV. The line energies are from \cite{1967RvMP...39...78B}. The abundance patterns for different types of white dwarfs used in these calculations  are summarized in Table \ref{tab:abund}.}

\end{minipage}
\end{table*}

\section{Discussion}

Results of  the previous section suggest that  absence  of a strong $\rm K_\alpha$ line of iron at 6.4 keV in the spectrum of a UCXB should be considered as an indication of a C/O/Ne white dwarfs whereas the presence of such a line points at a helium-rich donor.  There are, however, several factors which can modify this simple picture.

\subsection{Gravitational sedimentation of metals}

Gravitational settling of heavy elements  operates efficiently in the surface layers of white dwarfs, its time scale being considerably shorter than the evolutionary time scale of an isolated white  dwarf \citep[e.g.][]{1958QB843.W5S3.....,1986ApJS...61..197P}. The result of the gravitational sedimentation is that the outer layers of the white dwarf envelope  are depleted of metals on the time scales of $\sim 10^6$ yrs or shorter \citep{1986ApJS...61..197P,1992ApJS...82..505D}. This explains the purity of spectra in the majority of white dwarfs, which show only spectral lines of hydrogen and helium.

Extrapolating these results to the UCXB case, one may expect that their accretion disks should be devoid of elements other than the main constituent of the white dwarf (H/He or C/O). However, although the diffusion time scale in the outermost layers  is much shorter than the cooling time of an isolated white dwarf, the time scale on which the deeper layers of the white dwarf envelope are depleted of metals is rather long, much longer  than the accretion time scale. As a result, the outer layers of the white dwarf envelope are removed faster than they are depleted of metals. The time dependent calculations of gravitational settling by \citeauthor{1992ApJS...82..505D} show that time, required to deplete  the outer $\sim 10^{-3}$ (by mass) of the white dwarf exceeds $\sim 10^8$ yrs. On the other hand, for the accretion rate of $\sim 10^{-9}$ M$_\odot$/yr, the outer layer of mass $\Delta M\sim 10^{-3}$ M$_\odot$ is removed within $\Delta t\sim 10^6$ yrs.  Therefore we do not expect that gravitational settling of heavy elements is an important factor determining  the composition of accretion disks in UCXBs. This conclusion is confirmed by detection of iron lines  in the spectra of UCXBs with He-rich donors \citep{2000ApJS..131..571A,2004A&A...418.1061B}.

\subsection{Ionization state of the accretion disk}
\label{sec:ioniz}

The material on the surface of the accretion disk may be ionized due to internal heating as well as due to irradiation. This can significantly modify the shape of the  reflected emission and strengths of fluorescent lines. For example, if carbon and oxygen are fully ionized, they do not contribute to the ionization cross-section and the fluorescent line of iron will have its nominal, $\sim$solar abundance  strength even for the C/O-dominated disk. On the other hand, partial ionization of oxygen would lead to appearance of much stronger lines of OVII--OVIII, but would not eliminate the effect of screening of elements with higher charge $Z$. Correspondingly,  the iron line would remain suppressed in this case.

Illumination of the accretion disk by X-ray photons produced in the vicinity of the compact object (e.g. in the boundary layer) results in the appearance of the  ionized skin -- a thin surface layer of highly ionized material \citep{2000ApJ...537..833N}. Beneath this layer, the disk remains in the ionization state determined by heating due to viscous dissipation. The Thomson optical depth $\tau$ of the ionized skin is determined by the  "gravity parameter",
\begin{equation}
A=\frac{H}{R} \frac{L_{\rm{Edd}}}{L\rm_x \cos\theta}
\label{eq:gravpar}
\end{equation}
where $H/R$ is the aspect ratio of the Shakura Sunyaev accretion disk, $L\rm_x$ is the luminosity of the irradiating source, $L\rm_{Edd}$ -- Eddington luminosity for a neutron star and $\theta$ -- the angle of the incident radiation with respect to the normal to the disk surface. This parameter characterizes the strength of gravity relative to that of radiation pressure.  For typical parameters of UCXBs, $H/R\sim {\rm few}\times 10^{-2}$, $L_{\rm{x}}/L\rm_{Edd}\sim 10^{-2}-10^{-1}$ and assuming $\cos\theta\sim 0.1$ we obtain $A\sim 3-30$. According to \citeauthor{2000ApJ...537..833N}, this value corresponds to low to moderate illumination, in which case the Thomson optical depth of the ionized skin is small  and it does not significantly distort the spectrum of radiation reflected off the inner layers of the disk. The ionization state of these layers is determined by the viscous dissipation.

The effective temperature of the Shakura-Sunyaev disk due to viscous dissipation \citep{1973A&A....24..337S} is
\begin{equation}
{T_{\rm{eff}} = {\left( {\frac{{3GM\dot M}}{{8\pi \sigma {r_{\rm{o}}}^3}}} \right)^{1/4}}{\left( {\frac{{{r\rm_o}}}{r}} \right)^{3/4}}{\left( {1 - \sqrt {{r\rm_o}/r} } \right)^{1/4}}}
\end{equation}
where $\sigma  = 5.67\cdot 10^{-5}{\rm{erg}}\,{\rm{c}}{{\rm{m}}^{ - 2}}\,{{\rm{s}}^{ - 1}}\,{{\rm{K}}^{ - 4}}$
is the Stefan-Boltzmann constant, $M$ is the mass of the accretor ($ \sim 1.4\,{M_ \odot }$ for a neutron star), $\dot M$ the mass accretion rate and ${r\rm_o}$ is the inner radius of the disk, which in the case of the soft state of a neutron star is equal to its radius.   For the mass accretion rate in the $\dot{M}\sim 10^{-10}-10^{-8}~M_\odot$/yr range (corresponding to $L\rm_X\sim 10^{36}-10^{38}\rm erg\,s^{-1}$) the maximum disk temperature ranges from   $\sim 2\cdot 10^6$   to $\sim 8\cdot 10^6$ K. Obviously, hydrogen and helium are fully ionized throughout the most of the accretion disk at any value of the accretion rate relevant to UCXBs (as they should be, in order for accretion to proceed in the high viscosity regime). Iron is virtually never fully ionized, its ionization state being below FeXXIII in the entire temperature range of interest, $T\rm_{eff}\la 8\cdot 10^6$ K.
The ionization states of carbon, oxygen and neon, however,  depend critically on the mass accretion rate.  Indeed, for plasma in the collisional ionization equilibrium, in coronal approximation, 90\% of carbon (oxygen) is in the fully ionized state at temperatures of $\sim 1.6\cdot 10^{6}$ ($\sim 3.5\cdot 10^{6}$)  K \citep{1982ApJS...48...95S}. Therefore, even at the moderate luminosity of $\sim 10^{37}$ erg/s, carbon is expected to be ionized in a significant part of the inner disk, out to $r\sim 10 r\rm_o$. Oxygen and neon, on the other hand, are fully ionized in the inner disk at high luminosities, $\sim 10^{38}$ erg/s, but remain  only partially ionized at moderate and low luminosities,  $L\rm_X\sim 10^{36}-10^{37}$ erg/s.

Thus, one should expect luminosity dependence of the iron line strength in UCXBs. At low and moderate luminosities, it is determined by the  chemical composition of the accreting material, as described in the previous section. In particular, it remains at the solar value in systems with He-rich donors and is  suppressed in the case of a C/O/Ne donor. As luminosity\footnote{ Determination of the exact value of luminosity at which the transition happens, requires detailed calculations of the ionization state of the material at the surface of the  accretion disk and is beyond the scope of this paper.} approaches $\sim 10^{38}$ erg/s, the iron line should recover its nominal strength,  determined by the abundance of iron in the accreting material.

\subsection{Reflection from the white dwarf surface}

For a binary system consisting of a $\rm 1.4 M_{\odot}$ NS accretor and a $\rm 0.7 M_{\odot}$ WD donor, the Roche lobe of the WD subtends a solid angle of $\rm \sim 0.3\,sr$ as viewed from the neutron star. In the idealized lamppost geometry the solid angle of the  accretion disk is $\Omega \sim 2\pi$. Therefore, the donor star does not contribute significantly to the total reflected emission. However, the lamppost geometry is obviously a too crude approximation. In more realistic geometries, taking into account the geometry and emission diagram of the primary emission,  the role of reflection from the donor star may no longer be negligible.

For a $T\rm_{eff}\sim 10^4$ K white dwarf, helium is expected to be partly or fully ionized, whereas carbon, oxygen and iron should be only partly ionized \citep{1992ApJS...82..505D}. Therefore,  for a He white dwarf, the fluorescent line of iron should have its nominal strength. For a C/O white dwarf the iron line should be suppressed due to screening by carbon and oxygen, as described above. Depending on the exact value of the temperature in the white dwarf photosphere, carbon and oxygen may be in the medium to high ionization state. In the latter case, one may expect appearance of strong resonant lines of their highly ionized species. Among these, of special interest would be lines of oxygen.

\subsection{Iron abundance}

In order to determine the iron abundance, we relied on the assumption that the total mass of elements, heavier than O/Ne, in the donor star, does not change. This  was  motivated by the fact that nuclear reaction chains do not involve heavy elements. Moreover, the heavier elements were fixed at their solar mass fractions, in order to make comparisons between spectra originating in H-rich and H-poor atmospheres. These assumptions, along with the requirement to conserve the total number of nucleons, effectively  limited the  maximal C,O,Ne/Fe abundance ratios. This in turn limited  the minimal value  of the iron line EW  at  small but moderate values of $\sim\,$several eV (Fig. \ref{fig:co}, Table \ref{tab:ews}). However, if one allowed  further increase of the O/Fe abundance ratio, the equivalent width of the iron line can decrease to arbitrarily small numbers, in inverse proportion to this ratio, as illustrated by the Fig. \ref{fig:pz_vs_ab}.

\subsection{Comparison with observations}
\label{sec:obs}

Although we plan to perform detailed comparison of our predictions with observations of UCXBs with Chandra and XMM-Newton in a follow-up publication, we present some preliminary conclusions based on the published results. As already mentioned in the introduction, an iron line has been detected in the spectrum of UCXB 4U 1916-05 \citep{2000ApJS..131..571A,2004A&A...418.1061B}, which is believed to have a He-rich donor \citep{2006MNRAS.370..255N}. Similarly 4U 1820-30, for which \citet{1995ApJ...438..852B} and \citet{2003ApJ...595.1077C} have predicted a He-rich donor, has also been fitted with an iron $ \rm K_{\alpha}$ emission line \citep{2010ApJ...720..205C}. On the other hand, objects 4U 0614+091 and 4U 1543-624, that show signatures of C/O-rich material at optical wavelengths \citep{2004MNRAS.348L...7N,2006MNRAS.370..255N}, seem to have very faint, if any, iron $ \rm K_{\alpha}$ emission \citep{2010MNRAS.407L..11M,2011MNRAS.412L..11M}.

However, the picture is far from being complete and conclusive, with different analysis and different observations giving sometimes contradicting results. For example, \citet{2010A&A...522A..96N} found an iron line in the XMM spectrum of 4U 1543-624 with an EW of $\rm\sim 30\,eV$, whereas \citet{2012A&A...539A..32C} did not detect an iron line in the  of XMM spectra of 4U 1820-30. In addition \citet{2010A&A...514A..65K} interpret the observed bursting activity of 4U 0614+091, by considering a He-rich or hybrid donor, instead of a C/O-rich one. We plan to conduct systematic investigation of XMM-Newton and Chandra observations of UCXBs in a separate paper.

\section{Summary and conclusions}

We have shown that non-solar composition of the donor star in ultra-compact X-ray binaries may have a dramatic effect on the reflected spectral component in UCXBs, significantly modifying the strength of fluorescent lines of various elements.

To identify the main trends, we considered an idealized  case of an optically thick slab of neutral material with significantly non-solar abundances. We considered two abundance patterns, corresponding to a He and C/O white dwarf.
We found that from the observational point of view, the  non-solar composition  of the reprocessing material most pronouncedly affects  the strength of the fluorescent line of iron at 6.4 keV. Although increase of the carbon and oxygen abundances does lead to some increase of the strengths of corresponding fluorescent lines, their equivalent widths saturate at the (sub-)eV level due to the effect of self-screening. On the other hand, the equivalent width of iron decreases nearly in inverse proportion to the C/O abundance and the line is expected to be  significantly  fainter for the chemical composition of the C/O white dwarf.  This is caused by the screening of the presence of heavy elements by oxygen. In C/O-dominated material, the dominant interaction process for a $E\ga 7$ keV photon is absorption by oxygen rather than by iron, contrary to the case of solar composition. Screening by helium is significantly less important, due to its lower ionization threshold. Moreover, helium is expected to be fully ionized in the accretion disks of UCXBs. Consequently, in the case of He-rich reprocessing material,  fluorescent lines of major elements  are near their nominal, solar abundance strength. Thus, the equivalent width of the fluorescent line of iron can be used for diagnostics of the donor star in UCXBs by means of  X-ray spectroscopy.

In the realistic case of reflection in UCXBs, gravitational settling of elements and ionization of the disk material may, potentially, complicate and modify this picture. Simple comparison of the diffusion time in white dwarf envelope and the accretion time scale suggests that gravitational settling  is not fast enough to deplete iron and other heavy elements in the accreting material.  However, a more accurate consideration of the physical state of a Roche-lobe filling, mass losing white dwarf may still be needed for the final conclusion.

Ionization of the disk material at high mass accretion rate may lead to luminosity dependence of the discussed effects. In particular, as oxygen in the inner parts of the C/O-dominated disk becomes fully ionized at high mass accretion rate, its screening effect vanishes and the iron line should be restored to its nominal value, determined by the abundance of iron in the accretion disk. Comparison of ionization curves with the effective temperature distribution in the Shakura-Sunyaev accretion disk suggests that it should happen at the $\log(L\rm_X)\sim 37.5-38$ level. At lower luminosities, $\log(L\rm_X)\la 37$, oxygen in the inner disk is in the low- to medium ionization state and the idealized picture outlined above holds, at least qualitatively, and the strength of the 6.4 keV iron line may be used for diagnostics of the nature of the donor star. In particular, its absence points at the C/O or O/Ne white dwarf, while its presence suggests a He-rich donor.

\section*{Acknowledgements}
LB acknowledges support by the National Science Foundation under grants PHY
11-25915 and AST 11-09174.

\bibliographystyle{mn2e}
\bibliography{X-ray_diag_rev}

\appendix

\section[]{Single scattering approximation}
{We consider  reflection from a semi-infinite atmosphere in the single scattering approximation.  The spectral intensity of reflected emission $S\rm_{refl}$ $\rm(phot\,sec^{-1}\,cm^{-2}\,keV^{-1}\,sr^{-1})$ is given by the following expression.
\begin{eqnarray}\nonumber
{\delta S_{\rm{refl}}(E,{\vec n_{\rm{out}}})} = {S_{\rm{pr}}(E,{\vec n_{ \rm{in}}})}\,\delta\Omega_{\rm{in}}\\
\int\limits_0^\infty  dz\sec {\theta_{ \rm{in}}}\, \nonumber
{e^{ - n{\sigma_{ \rm{tot}}(E)}z\sec {\theta_{\rm{in}}}}}\,n{\sigma_{ \rm{sc}}(E)}{P_{\rm{sc}}}({\vec n_{ \rm{in}}},{\vec n_{ \rm{out}}})\\
{e^{ - n{\sigma_{ \rm{tot}}(E)}z\sec {\theta_{ \rm{out}}}}}
\label{eq:srefl}
\end{eqnarray}
where $S\rm_{pr}(E,\vec n)$ is the spectral intensity of the primary radiation, axis $z$ is normal to the surface of the atmosphere and is directed inwards, $\delta\Omega\rm_{in}$ is an infinitesimal solid angle around the direction of incidence $\vec n \rm_{in}$,  ${\theta \rm_{in}}$ is its polar angle  with respect to the axis $z$. The spectral intensity of the reflected emission is computed at the direction $\vec n \rm_{out}$, which polar angle is ${\theta \rm_{out}}$. ${P\rm_{sc}}({\vec n \rm_{in}},{\vec n \rm_{out}})$ is the probability that the photon, entering the medium from the direction $\vec n \rm_{in}$ is scattered in the  direction $\vec n \rm_{out}$, it is normalized so that:
\begin{equation}
\int {P\rm_{sc}}({\vec n \rm_{in}},{\vec n \rm_{out}}) \, d\Omega\rm_{out} = 1
\end{equation}
The $n$ is the density of the material and ${\sigma \rm_{tot}} = {\sigma \rm_{abs}} + {\sigma \rm_{sc}}$ is the total cross section. The   absorption cross section due to photoionization ${\sigma \rm_{abs}}$ is given by the following expression
\begin{equation}
 {\sigma _{{\rm{abs}}}}(E) = \sum\limits_{Z = 1}^{30} {{A_Z}{\sigma _Z}(E)}
\end{equation}
where we account for all elements from $Z=1$ to $Z=30$,  $A\rm_Z$ is the abundance of element Z by the particle number, and ${\sigma \rm_Z}$ is the photoionization cross-section for all shells of element Z. It is  calculated using the second version of the \citet{1996ApJ...465..487V} subroutine. ${\sigma \rm_{sc}}$ is the scattering cross section per hydrogen atom, given by
\begin{equation}
{\sigma _{{\rm{sc}}}}(E) = \sum\limits_{Z = 1}^{30} {Z{A_{\rm{Z}}}{\sigma _{\rm{T}}}}
\end{equation}
where $\sigma\rm_T$ is the Thomson cross section. Note that we consider Compton scattering in the low energy limit and ignore change of the photon frequency during scattering  in deriving eq.(\ref{eq:srefl}).

For a semi-infinite atmosphere, the reflected spectrum does not depend on the density of the material, only on its chemical composition.

We assume for simplicity that the energy and angular dependencies of the primary radiation  can be factorized:
\begin{equation}
{S_{{\rm{pr}}}}\left( {E,\vec n} \right) = {S_{\rm{0}}}(E)\,{P_{{\rm{pr}}}}(\vec n)
\end{equation}
where the ${P\rm_{pr}}({\vec n\rm_{in}})$ describes the angular distribution of the primary radiation and is normalized so that
\begin{equation}
\int {P\rm_{pr}}({\vec n}) \, d\Omega = 1
\end{equation}
and $S\rm_0(E)$ is proportional to the total luminosity of the primary emission and has units of $\rm phot\,sec^{-1}\,cm^{-2}\,keV^{-1}$.
We evaluate the reflected emission within solid angle $\Delta\Omega\rm_{out}$ around direction of interest $\vec n \rm_{out}$:
\begin{equation}
{\tilde S_{{\rm{refl}}}}(E) = {\int_{\Delta {\Omega _{{\rm{out}}}}} S _{{\rm{refl}}}}\left( {E,{{\vec n}_{{\rm{out}}}}} \right)d\Omega
\end{equation}
Integrating eq. A1 over ingoing and outgoing directions and from z=0 to $z= \infty $ we obtain the following expression for  the reflected continuum $\tilde S\rm_{refl}$
\begin{equation}
{\tilde S_{{\rm{refl}}}}(E) = {S_{\rm{0}}}(E)\,\frac{{{\sigma _{{\rm{sc}}}}(E)}}{{{\sigma _{{\rm{sc}}}}(E) + {\sigma _{{\rm{abs}}}}(E)}}\,{R_{{\rm{refl}}}}
\label{eq:srefl1}
\end{equation}
where  the factor $R\rm_{refl}$ accounts for the geometry of the problem and is given by
\begin{eqnarray}\nonumber
{R\rm_{refl}} &=& \int_{\Delta\Omega\rm_{in}} {d{\Omega \rm_{in}}{P\rm_{pr}}({{\vec n}\rm_{in}})} \int_{\Delta\Omega\rm_{out}} {d{\Omega \rm_{out}}}\\
 &&\sec {\theta \rm_{in}}\,{\left( {\sec {\theta \rm_{in}} + \sec {\theta \rm_{out}}} \right)^{ - 1}}{P\rm_{sc}}({\vec n\rm_{in}},{\vec n\rm_{out}}),
\label{eq:rrefl}
\end{eqnarray}
 For Thomson scattering ${P\rm_{sc}}({\vec n\rm_{in}},{\vec n\rm_{out}})$ depends only on the scattering angle $\theta\rm_{sc}$ and  is given by the standard Rayleigh formula
\begin{equation}
{P\rm_{sc}}(\theta\rm_{sc}) = \frac{3}{{8\pi }}\frac{{\left( {1 + {{\cos }^2}{\theta \rm_{sc}}} \right)}}{2},
\end{equation}
Ignoring angular dependences,
\begin{equation}
R\rm_{refl}\propto \frac{\Delta\Omega\rm_{in}}{4\pi} \, \frac{\Delta\Omega\rm_{out}}{4\pi}
\end{equation}
For the case of isotropic incident radiation and reflected emission integrated over all outgoing angles (the geometry assumed in Fig. \ref{fig:co}, \ref{fig:he})  eq. A9 yields $ \rm R\rm_{refl}=1/8$.

Using the same approach and  taking into account the fluorescent yield, we can compute the  fluorescent line flux  $F\rm_{line}$ ($\rm phot\,sec^{-1}\,cm^{-2}\,sr$):
\begin{eqnarray}\nonumber
{\delta F_{\rm{line}}({\vec n_{ \rm{out}}})}=\delta\Omega_{\rm{in}} \int\limits_{{E_{\rm{K}}}}^\infty {dE}\,S_{\rm{pr}}(E,{\vec n_{\rm{in}}})\\
\int\limits_0^\infty  dz\sec {\theta_{ \rm{in}}}\, {e^{ - n{\sigma_{ \rm{tot}}}(E)z\sec {\theta_{ \rm{in}}}}}{A_{\rm{Z}}}n\,{\sigma _{\rm{K,Z}}}{Y_{\rm{Z}}}{P_{\rm{line}}}({\vec n_{\rm{out}}})\nonumber\\
{e^{ - n{\sigma_{ \rm{tot}}}({E_{\rm{line}}})z\sec {\theta_{ \rm{out}}}}}
\label{eq:Fline}
\end{eqnarray}
In the above expression, $\sigma \rm_{K,Z}$, is the K-shell absorption cross section of element Z and $Y\rm_Z$ is the fluorescence yield of its  $\rm K_\alpha$  line, $E\rm_{line}$ is the $\rm K_\alpha$ line energy and ${E\rm_K}$ is the energy of the K-edge. ${P\rm_{line}}(\vec n\rm_{out})$ is the angular distribution of the fluorescent emission, assumed to be isotropic ($P\rm_{line}(\vec n\rm_{out})=1/4\pi$).
Integrating over all angles and over $z = 0 \to \infty $ we obtain
\begin{equation}
{\tilde{F}\rm_{line}} = \int\limits_{{E\rm_K}}^\infty  S_{0}(E)G(E)dE
\label{eq:fline1}
\end{equation}
where as before,  tilde denotes integration over solid angle $\Delta\Omega\rm_{out}$ and $G(E)$ is given by the following integral
\begin{eqnarray}\nonumber
G(E) &=& \int_{\Delta\Omega\rm_{in}} {d{\Omega \rm_{in}}}{P\rm_{pr}}({\theta \rm_{in}}) \int_{\Delta\Omega\rm_{out}} {d{\Omega \rm_{out}}}{P\rm_{line}}({\vec n\rm_{out}}) \\
&& \left( {\frac{{{{Y\rm_Z A\rm_Z}\,\sigma \rm_{K,Z}}(E)\sec {\theta \rm_{in}}}}{{{\sigma \rm_{tot}}(E)\sec {\theta \rm_{in}} + {\sigma \rm_{tot}}({E\rm_{line}})\sec {\theta \rm_{out}}}}} \right)
\label{eq:ge}
\end{eqnarray}
A similar approach was used by \cite{2008MNRAS.385..719C} in calculating the Earth albedo. Similarly to $R\rm_{refl}$, for the geometry of Fig. \ref{fig:co}, \ref{fig:he} (semi-infinite slab illuminated by isotropic incident radiation, reflected emission integrated over all outgoing angles),  eq. \ref{eq:ge} can be integrated  analytically to give:
\begin{eqnarray}
G(E) = {Y\rm_Z}{A\rm_Z}\,{\sigma \rm_{K,Z}}(E)\,\frac{{G1}}{{G2}}
\end{eqnarray}
Where G1 is given by,
\begin{eqnarray}
G1 = {\sigma \rm_{tot}}({E\rm_{line}})[{\sigma \rm_{tot}}(E) + {\sigma \rm_{tot}}({E\rm_{line}})\ln [{\sigma \rm_{tot}}({E\rm_{line}})]] \nonumber \\
+ [{\sigma \rm_{tot}}{(E)^2} - {\sigma \rm_{tot}}{({E\rm_{line}})^2}]\ln [{\sigma \rm_{tot}}(E) + {\sigma \rm_{tot}}({E\rm_{line}})]\nonumber\\
 - {\sigma \rm_{tot}}{(E)^2}\ln [{\sigma \rm_{tot}}(E)]
\end{eqnarray}
and G2 is given by,
\begin{eqnarray}
G2 = 8\,{\sigma \rm_{tot}}{(E)^{2}}{\sigma \rm_{tot}}({E\rm_{line}})
\end{eqnarray}

To characterize  the strength of emission lines we evaluate their equivalent widths  with respect to the total continuum emitted within the solid angle $\Delta\Omega\rm_{out}$ around the direction of interest.
\begin{equation}
{EW = \frac{{{\tilde F\rm_{line}}}}{{{\tilde S\rm_{tot}}({E\rm_{line}})}}},
\label{eq:ew}
\end{equation}
The total continuum includes both the reflected continuum, given by eq. (\ref{eq:srefl1}) and the fraction of the primary continuum emitted  in the solid angle $\Delta\Omega\rm_{out}$
\begin{equation}
{\tilde S_{{\rm{tot}}}}(E) = {\tilde S_{{\rm{refl}}}}(E) + {\tilde S_{{\rm{pr}}}}(E)
\label{eq:stot}
\end{equation}
where, similarly to $\tilde S\rm_{refl}(E)$,
\begin{equation}
{\tilde S_{{\rm{pr}}}}(E) = {\int_{\Delta {\Omega _{{\rm{out}}}}} S _{{\rm{pr}}}}\left( {E,{{\vec n}_{{\rm{out}}}}} \right)d{\Omega _{{\rm{out}}}}
\label{eq:sin}
\end{equation}

Thus, using equations \ref{eq:srefl1} and \ref{eq:rrefl} for the reflected continuum,  eqs. \ref{eq:fline1} and \ref{eq:ge} for the fluorescent line flux, the  equivalent width of the fluorescent line can be computed from eqs. \ref{eq:ew}--\ref{eq:sin}.  Comparison with Monte-Carlo calculations show that the single scattering approximation works nearly perfectly at low energies, $E\la 2$ keV,  where absorption dominates scattering. At higher energies,  multiple scattering becomes more important, resulting in a $\sim 10\%$ offset  between analytical and Monte-Carlo results for the iron $\rm K_{\alpha}$ line. This is further illustrated by  Figs. \ref{fig:co} and \ref{fig:he} showing dependence of equivalent widths of various fluorescent lines on the chemical abundances  computed in single scattering approximation and using the  Monte-Carlo code.

\label{lastpage}
\end{document}